\documentclass[12pt]{article}
\usepackage{epsfig}
\begin{document}
\setlength{\baselineskip}{0.20in}
\newcommand{\nc}{\newcommand}
\newcommand{\beq}{\begin{equation}}
\newcommand{\eeq}{\end{equation}}
\newcommand{\be}{\begin{eqnarray}}
\newcommand{\ee}{\end{eqnarray}}
\newcommand{\num}{\nu_\mu}
\newcommand{\nue}{\nu_e}
\newcommand{\nut}{\nu_\tau}
\newcommand{\nus}{\nu_s}
\newcommand{\mnus}{M_s}
\newcommand{\taus}{\tau_{\nu_s}}
\newcommand{\nnt}{n_{\nu_\tau}}
\newcommand{\rnt}{\rho_{\nu_\tau}}
\newcommand{\mnt}{m_{\nu_\tau}}
\newcommand{\tnt}{\tau_{\nu_\tau}}
\newcommand{\bi}{\bibitem}
\newcommand{\rar}{\rightarrow}
\newcommand{\lar}{\leftarrow}
\newcommand{\lrar}{\leftrightarrow}
\newcommand{\dm}{\delta m^2}
\newcommand{\so}{\, \mbox{sin}\Omega}
\newcommand{\co}{\, \mbox{cos}\Omega}
\newcommand{\sotil}{\, \mbox{sin}\tilde\Omega}
\newcommand{\cotil}{\, \mbox{cos}\tilde\Omega}
\makeatletter
\def\alt{\mathrel{\mathpalette\vereq<}}
\def\vereq#1#2{\lower3pt\vbox{\baselineskip1.5pt \lineskip1.5pt
\ialign{$\m@th#1\hfill##\hfil$\crcr#2\crcr\sim\crcr}}}
\def\agt{\mathrel{\mathpalette\vereq>}}

\newcommand{\eq}{{\rm eq}}
\newcommand{\tot}{{\rm tot}}
\newcommand{\M}{{\rm M}}
\newcommand{\coll}{{\rm coll}}
\newcommand{\ann}{{\rm ann}}
\makeatother

\begin{center}
\vglue .06in
{\Large \bf Matter-antimatter domains in the universe}\\
{\it Invited Talk at  EuroConference on Frontiers in Particle Astrophysics
and
    Cosmology, San Feliu de Guixols, Spain, 30 September -5 October
2000}\\

\bigskip
{\bf A.D. Dolgov
\footnote{Also: ITEP, Bol. Cheremushkinskaya 25, Moscow 113259, Russia.}
\\
{\it{INFN section of Ferrara\\
Via del Paradiso 12,
44100 Ferrara, Italy}
}}
\\[.40in]
\end{center}

\begin{abstract}
A possible existence of cosmologically large domains of antimatter 
or astronomical ``anti-objects'' is discussed. A brief review of different 
scenarios of baryogenesis predicting a noticeable amount of antimatter 
is given. Though both theory and observations indicate that the universe 
is most possibly uniformly charge asymmetric without any noticeable 
amount of antimatter, several natural scenarios are possible that allow 
for cosmologically (astronomically) interesting objects in close vicinity 
to us. The latter may be discovered by observation of cosmic ray
antinuclei.

\end{abstract}

\section{INTRODUCTION}

It is well known that the world of elementary particles is doubled - for
each particle there exists an antiparticle with the same mass, $m=\bar m$,
and life-time (if unstable), $\tau = \bar \tau$, but with opposite signs
of charges associated with any conserved vector current, $Q_v =-\bar Q_v$.
Scalar or tensor charges have the same signs for particles and 
antiparticles. These are the results of invariance of any
``normal'' theory with respect to 
combined transformation of charge conjugation (C), mirror reflection (P),
and time-reversal (T), that follows from famous CPT-theorem. Despite this
symmetry the universe is seemingly populated only by particles, at least
in
our neighborhood. However, neither C-transformation, nor combined
CP-parity
one are symmetries of the theory. Hence properties of particles and 
antiparticles (or even mirror
reflected antiparticles) are slightly different. This small difference 
together with a possible non-conservation of baryonic charge and a 
deviation from thermal equilibrium in the course of cosmological expansion
created an overwhelming dominance of matter with respect to 
antimatter~\cite{sakharov67}. In the simplest versions of the 
baryogenesis scenario (for the review see~\cite{dolgov92})
the cosmological baryon asymmetry
\be
\beta = {(N_B - N_{\bar B}) \over N_\gamma}
\label{beta}
\ee
is a universal constant and there is no room for astronomically large
antimatter domains. However in more complicated models the baryon
asymmetry may be non-uniform, $\beta = \beta (x)$, creating the so called
isocurvature perturbations and, moreover, it might change sign so that
large parts of the universe would be built of antimatter. Unfortunately
there is no definite theory so neither the characteristic size of the
domains nor the distance to them can be predicted with any certainty. 
Still the idea that there can exist cosmic domains of antimatter
or even that the universe is globally charge symmetric, is very
attractive and the question if there is any cosmological antimatter 
remains with us for several decades after it was discovered that 
antimatter in principle exists.

\section{OBSERVATIONAL LIMITS}

There are two possible ways to search for cosmic antimatter. First, one
may look for it in the cosmic rays. There is a small amount of 
energetic antiprotons at the level of $10^{-4}$ with respect to 
protons. It can be explained by the secondary production of $\bar p$ 
in proton collisions. It is very difficult to create antinuclei in 
proton-proton or proton-nucleus collisions. Their registration in the
cosmic rays would be a strong indication to the existence of cosmic
antimatter. With a small probability antideuterons may be produced in
annihilation of neutralino dark matter~\cite{donato99} but secondary
production of heavier elements is practically impossible. The recent
preliminary data obtained by Alpha Magnetic
Spectrometer~\cite{battiston99}
give the following upper limits for the flux of antihelium and heavier 
elements;
\be
\Phi\,(^4\bar He) /\Phi\,(^4He) < 2\cdot 10^{-6} \nonumber\\
\Phi\,(\bar Z>2) /\Phi\,(Z>2) < 2\cdot 10^{-5}
\label{fluxes}
\ee  
More activity is expected in this area and more restrictive bounds 
will be obtained or maybe even a discovery of antinuclei will be made 
in the nearest future.

In the absence of definitive theoretical predictions we will discuss
logical possibilities for the behavior of $\beta$ and reasonable
scenarios for realization of one or other logical option. The simplest
(and dullest) model is that $\beta = const$. In this case all the 
universe is homogeneously populated with baryonic matter and the 
search for antimatter would be fruitless. Another option is that
the baryon asymmetry is a varying function of space points, 
$\beta = \beta (x)$ but $\beta > 0$ everywhere. In this model 
isocurvature density perturbations created by the different chemical
(baryonic) content might be quite essential but still no antimatter 
domains would exist. If $\beta \neq const$, then it is quite natural
that it might have somewhere negative values so these parts of the
universe would be filled by antimatter. A particular case of globally
symmetric universe, when there is an equal amount of matter and
antimatter:
\be
\int \beta dV =0
\label{intbeta}
\ee
is aesthetically attractive. However there exist scenarios when the
universe is dominated by matter while relatively small amount of 
antimatter in the form of gas clouds or separate anti-stars or even
rare anti-galaxies is possible~\cite{dolgov93}. Such small admixtures
of antimatter may be relatively close to us. 

There are several ways in which cosmic antimatter may become visible.
First, $p\,\bar p$-annihilation into 
$\pi$-mesons would produce $\sim 100$ MeV
cosmic $\gamma$-rays coming from the decay $\pi^0 \rar 2\gamma$ or
from annihilation of energetic positrons from the sequence of the decays 
$\pi^+\rar \mu^+ \rar e^+$. 

The old analysis of reference~\cite{steigman76}
permits to conclude that if antimatter in large amount exists, it should 
be at least at the distance 10 Mpc. This conclusion was revised recently
in
ref.~\cite{cohen98} where a much stronger limit was obtained that
a possible anti-world should be very far away from us at a distance larger 
than $\sim $ Gpc. This result was obtained under assumption of 
baryo-symmetric universe. It was also assumed that matter and antimatter
domains are in close contact. The last assumption is justified by the 
smoothness of the cosmic microwave background radiation (CMBR). The
latter demands that there cannot be density contrast above $10^{-4}$
at the scale larger than 15 Mpc. On the other hand, according to the
calculations of ref.~\cite{cohen98} the photon diffusion would drag matter
to a larger distance. Hence even if the matter-antimatter domains
were originally spatially separated (but less than for 15 Mpc) they
would come to a close contact due to the photon drag. After annihilation
would start its products would efficiently carry energy far away from the
annihilation region because 
a bulk of the produced particles are energetic
electrons or positrons and neutrinos whose mean free path is large in
comparison with the annihilation region. Hence contrary to naive
expectations, the energy and pressure density in the region of 
annihilation becomes smaller and it would increase the diffusion of 
matter and antimatter towards each other and amplify the efficiency
of annihilation. Such annihilation after hydrogen recombination would
be a source of a very strong gamma ray background. Non-observation 
of this background permits to obtain a very restrictive 
limit that any abundant cosmic antimatter should be near or beyond the
horizon. However this result is not valid for the case of isocurvature
fluctuations (see discussion below).

Another possible effect of matter-antimatter annihilation is a distortion
of energy spectrum of CMBR if the annihilation takes place before
recombination. This could happen for large size domains. This effect
was considered in refs.~\cite{kinney97,cohen97}. At the present time
this phenomenon does not permit to obtain any interesting limit. 

The limits discussed above should be very much weaker in the case of
isocurvature fluctuations. In this case initial density contrast was 
zero and started to rise only relatively late. Baryon (antibaryon) rich
regions cooled faster so photons would diffuse there from hotter baryon
poor regions and this diffusion would drag matter and antimatter 
{\it away} so baryons and antibaryons would go out of contact. This
picture 
is opposite to the one considered above. In this model annihilation would
be very weak and one would expect the universe consisting of possible
large
matter and antimatter domains separated by relatively narrow 
baryon(antibaryon) voids. If the angular size of this voids is
sufficiently
small then the limits from angular fluctuations of CMBR are not applicable
and we may expect to have antimatter domains almost at hand.

\section{THEORETICAL MODELS}	

As has been already mentioned for generation of baryon asymmetry three
principles of baryogenesis~\cite{sakharov67} should be fulfilled:
\begin{enumerate}
\item{}
Non-conservation of baryonic charge.
\item{}
Breaking of C and CP invariance.
\item{}
Deviation from thermal equilibrium.
\end{enumerate}
There are several workable scenarios of baryogenesis. They all are
based on the assumption of explicit C and CP violation and give 
$\beta = const$ and no cosmic antimatter. However if charge symmetry
is broken spontaneously~\cite{lee73}, then in different CP-domains
the universe would be either baryonic or 
anti-baryonic~\cite{brown79}. The size of these
domains may be cosmologically large if after their formation the universe
passed through a period of exponential expansion
(inflation)~\cite{sato81}.
A review of the earlier ideas on the subject can be found
in~\cite{stecker85}. 
If C and CP are indeed broken spontaneously, then the universe should be
be globally charge symmetric with equal number of baryonic and 
anti-baryonic domains separated by domain wall with enormous mass. Such
domain walls would create serious cosmological
problems~\cite{zeldovich74}.
So either the size of the domains should be larger than the present day 
horizon and the antimatter would be unobservable or there should exist
a mechanism of domain wall destruction maintaining 
homogeneity and isotropy of the universe. However even if all this works,
the discussed mechanism would create matter-antimatter domains in close
contact with each other and with vanishingly small isocurvature density 
perturbations (because the baryon asymmetries in different domains have
differ by sign but have equal magnitude). This corresponds to the case
considered in ref.~\cite{cohen98} and, as we discussed above, permits
cosmic antimatter to be only very far away from us.

This pessimistic conclusion can be avoided in some more complicated
versions of baryogenesis scenario, in particular in the model of
spontaneous baryogenesis~\cite{cohen87} or in the model of scalar baryon
condensation~\cite{affleck85}. In the model of spontaneous baryogenesis
non-conservation of baryonic current is induced by a spontaneous breaking
of $U(1)$-symmetry associated with baryonic charge. The corresponding
(pseudo)goldstone field satisfies the equation of motion:
\be
D^2 \theta + U'(\theta) = f^{-2} \partial_\mu J^\mu_{bar}
\label{d2theta}
\ee
where $f$ is the scale of symmetry breaking and $J^\mu_{bar}$ is the
baryonic currents of fermions (quarks). In the case that the symmetry
is broken only spontaneously the potential $U(\theta)$ vanishes and the 
baryon charge density created by the evolution of the Goldstone field
$\theta$ is evidently given by
\be
B = f^2  \dot\theta_{in}
\label{dotthetha}
\ee
In the case that the potential $U$ is non-vanishing (pseudo-goldstone
field), e.g. $U(\theta) = m^2\theta^2$, the asymmetry is 
given by~\cite{dolgov95}
\be
B = f^2 \Gamma f^2 \theta_{in}^3,
\label{theta3}
\ee
where $\Gamma = g^2m/16\pi$ is the width of the decay of $\theta$ into
quarks.

One sees that the result for the asymmetry depends upon the initial value
of the field $\theta$ (or its derivative) and may have an {\it arbitrary 
sign}. The asymmetry can be generated without any explicit C and CP 
violation. The proper asymmetry between matter and antimatter is created
by charge asymmetric initial conditions and can be generated 
stochastically. This type of C, CP - violation can be called 
stochastic~\cite{dolgov92}. 
It is normally assumed that the initial conditions are formed
during inflation when a massless or light scalar field 
$\phi$ was infrared unstable and rose as
\be
\langle \phi^2 \rangle \sim H^3 t
\label{phi2}
\ee
reaching the average value $\sim H^4/m^2$~\cite{bunch78}. 
This models permits to have spatially varying $\beta (x)$ and even
domains of matter and antimatter at cosmologically large scales.
A more detailed discussion can be found in the 
refs.~\cite{dolgov92,dolgov93b,dolgov96}
or in a recent paper~\cite{khlopov00}. However this model suffers
from a large magnitude of isocurvature perturbation that may to be too 
high to be compatible with CMBR data~\cite{enqvist00}.

Similar picture of charge asymmetry generation may take place in the
Affleck-Dine model~\cite{affleck85}. In this model a condensate of 
a bosonic field with a non-zero baryonic charge might be formed along
a flat direction of the potential during inflationary stage. When
inflation
was over the field evolved down to the equilibrium point of the potential
and due to stochastic initial conditions started to ``rotate'' clockwise
or anti-clockwise, creating respectively baryons or antibaryons. In this
respect the model is similar to spontaneous baryogenesis discussed above,
however the asymmetry in this model might be much larger than in the
previously considered case. Moreover, the model may be combined with 
the explicit C and CP breaking in such a way that the bulk of the space 
would be filled with baryons with normal uniform density, 
and simultaneously relatively small bubbles with a much larger
asymmetry, $\beta$, may be formed~\cite{dolgov93}. 
Most probably such bubbles form primordial black holes
with log-normal mass distribution,
\be
dN/dM = \exp [-\gamma \ln^2 M/M_0 ]
\label{dndm}
\ee
with unknown constant parameters $\gamma$ and $M_0$. The remnants that
did not underwent gravitational collapse could be anti-stars,
clouds of antimatter, or even small isolated galaxies. Since the volume
occupied by this abnormal bubbles could be sufficiently small, their
existence is not forbidden and cosmological antimatter may be quite
close to us.

\section{CONCLUSION}

First and possibly rather gloomy conclusion is that most natural is to
expect that the universe is uniformly filled with baryons with a constant
asymmetry $\beta$ and there is no cosmologically noticeable antimatter. 
However the idea that the universe may be charge symmetric is quite 
attractive and there are natural theoretical frameworks for 
such cosmology. Simple versions of realization of charge symmetric world
comes into contradiction with observations of gamma-ray background and
with the data on CMBR. However some quite natural models with specific 
isocurvature density perturbations allow for large and relatively close
astronomical objects consisting of antimatter. They may be sources of
antinuclei in cosmic rays and the search for them could put stronger
limits on existence of cosmic antimatter or make a seminal discovery
of anti-worlds.

\end{document}